\documentclass[11pt]{article}
\usepackage{amsfonts}
\usepackage{amssymb}
\usepackage{amsmath}
\usepackage{graphicx}

\newcommand{\ket}[1]{|#1\rangle}

\def\01{\{0,1\}}

\def\x{\times}
\def\e{\varepsilon}

\def\ox{\otimes}

\def\ket#1{\mbox{$| #1 \rangle$}}

\def\ni{\noindent}

\def\ee{\vspace*{2mm}}
\def\loud#1{\noindent{\bf #1 }}
\def\GF{\mbox{\it GF\/}}

\def\sqx#1{#1{\x}#1}                       % m x m
\def\sqmtx#1#2{\mbox{$#2^{#1 \times #1}$}} % mxm sq mat over ring
\def\lcadd#1{\mbox{$C_{#1\ast}$}}          % left cont add operator
\def\rcadd#1{\mbox{$C_{\ast #1}$}}         % right cont add operator
\def\rlcadd#1{\mbox{$D_{#1\ast}$}}         % rev lt cont add operator
\def\rrcadd#1{\mbox{$D_{\ast #1}$}}        % rev rt cont add operator
\def\Z{\mathbb Z}                          % the integers
\def\trop#1{T_{#1}}                        % register-matrix transp
\def\trans{\mathsf{T}}                     % nice transpose symbol
\def\mathtab{\mbox{ }\mbox{ }\mbox{ }\mbox{ }\mbox{ }} % poor tabs

\newcommand{\op}[1]{\operatorname{#1}}
\def \qed {\hfill \rule{0.2cm}{0.2cm}\vspace{3mm}}

\newtheorem{theorem}{Theorem}

\newtheorem{definition}{Definition}[section]

\begin{document}

\title{\Large\bf
Sharp quantum vs.\ classical query complexity separations%
\footnote{Research partially supported by Canada's NSERC.}
}

\author{
J.~Niel de Beaudrap\thanks{Email: {\tt jd@cpsc.ucalgary.ca}}
\hspace{0.3in}
Richard Cleve\thanks{Email: {\tt cleve@cpsc.ucalgary.ca}}
\hspace{0.3in}
John Watrous\thanks{Email: {\tt jwatrous@cpsc.ucalgary.ca}}\\[3mm]
Department of Computer Science\\
University of Calgary\\
Calgary, Alberta, Canada T2N 1N4
}

\date{}

\maketitle

\thispagestyle{empty}

\begin{abstract}
We obtain the strongest separation between quantum and classical 
query complexity known to date---specifically, we define a black-box 
problem that requires exponentially many queries in the classical 
bounded-error case, but can be solved exactly in the quantum 
case with a single query (and a polynomial number of auxiliary 
operations).
The problem is simple to define and the quantum algorithm solving 
it is also simple when described in terms of certain quantum Fourier 
transforms (QFTs) that have natural properties with respect to the 
algebraic structures of finite fields.
These QFTs may be of independent interest, and we also investigate 
generalizations of them to noncommutative finite rings.
\end{abstract}

%============================================================================%

\section{Introduction}
\label{sec:introduction}

Shor's algorithm \cite{Shor97} for factoring integers in polynomial-time on
a 
quantum computer evolved from a series of quantum algorithms in the query
model.
This model appears to be useful for exploring the computational power of
quantum information.
In the query model, the input data is embodied in a black-box and the goal 
is to efficiently deduce some property of the black-box.
Efficiency is measured in terms of the number of queries made to the 
black-box.
A secondary measure of efficiency is also considered: the number of
auxiliary 
operations that must be performed to generate the input to the queries and 
process the output.
We will implicitly require that the number of auxiliary operations scales 
polynomially with the number of bits/qubits input to each query.

The first instance of a quantum algorithm outperforming a classical
algorithm
in the query model was due to Deutsch \cite{Deutsch85}, where a quantum
algorithm 
is able to solve a 2-bit query problem with one query (see also
\cite{CleveE+98}), 
whereas any classical algorithm for the problem requires two queries.
(A {\em $k$-bit\/} query is one that takes $k$ bits/qubits as input and 
returns $k$ bits/qubits as output.)
This was extended by Deutsch and Jozsa \cite{DeutschJ92}, who defined an 
$(n+1)$-bit query problem that can be solved exactly with one query by a 
quantum algorithm whereas it requires $\Omega(2^n)$ queries to solve
exactly
classically.
In spite of the apparent strength of this separation, the problem is only 
hard in the classical setting if the algorithm must be exact, meaning that 
no probability of error is tolerated.
A bounded-error algorithm is one that is allowed to err, provided that for 
any black-box instance its error probability is bounded below some constant
smaller than $1/2$.
There is a classical algorithm that solves the problem in \cite{DeutschJ92}
with bounded error using only $O(1)$ queries.

Subsequent work by Bernstein and Vazirani \cite{BernsteinV97} included an 
$(n+1)$-bit query problem that can be solved exactly with a quantum
algorithm 
making one query, whereas any bounded-error classical algorithm for it 
requires $n$ queries.
They also showed that a recursively defined version of this problem results
in a $\Theta(n)$-bit query problem whose exact quantum and bounded-error 
classical query complexities are $O(n \log n)$ and $n^{\Omega(\log n)}$, 
respectively.
This was improved by Simon \cite{Simon94}, who gives a fairly simple 
$O(n)$ vs.\ $\Omega(2^{n / 2})$ bounded-error quantum vs.\ bounded-error 
classical query separation.
Brassard and H{\o}yer \cite{BrassardH97} later showed that the problem
considered by Simon can in fact be solved exactly in the quantum
setting with $O(n)$ queries.

When cast in the query model, Shor's factoring algorithm can be viewed 
as an extension of Simon's work---it is a quantum algorithm that solves 
a $3n$-bit query problem with bounded-error with $O(1)$ quantum queries, 
while any classical algorithm for this problem requires 
$\Omega(2^{n/3} / \sqrt{n})$ queries (the lower bound is proved in 
\cite{Cleve00}).

What is the sharpest quantum vs.\ classical query complexity separation
possible?
For problems that can be solved exactly with a single quantum query, it
appears that the maximum classical bounded-error query complexity
previously-known for such a problem is $n$ \cite{BernsteinV97}.
However, if the efficiency and performance of the quantum algorithm are
relaxed to allow $O(1)$ queries and a bounded-error result then there is a
problem whose classical bounded-error query complexity is exponential
\cite{Shor97,Cleve00}.

Presently, we show that the best of the above two scenarios is possible 
by exhibiting a $2n$-bit query problem such that:
\begin{itemize}
\item
In the quantum setting, a single query suffices to solve the problem 
exactly.
Moreover, the auxiliary operations are very simple; they consist of 
$O(n)$ Hadamard gates followed by $O(n^2)$ classical gate operations 
that can occur after a measurement is made.
\item
In the classical setting, $\Omega(2^{n/2})$ queries to the black-box are 
necessary to solve the problem with bounded error.
\end{itemize}
The problem that achieves the above, which we call the {\em hidden 
linear structure\/} problem, is defined over the field $\GF(2^n)$ 
as follows.
Assume elements of the finite field $\GF(2^n)$ are identified with
strings in the set $\01^n$.
Let $\pi$ be an arbitrary permutation on $\GF(2^n)$ and let 
$r \in \GF(2^n)$.
Define the black-box $B$ as computing the mapping from 
$\GF(2^n) \times \GF(2^n)$ to itself defined as 
$B(x,y) = (x,\pi(y+sx))$.
The goal of the query problem is to determine the value of $s$.

It should be noted that this problem is related to, but different from,
the {\em hidden linear function\/} problem considered by Boneh and
Lipton \cite{BonehL95}.
In our problem, the linear structure occurs over the field $\GF(2^n)$ 
(and involves the multiplicative structure of $\GF(2^n)$), whereas for 
the hidden linear function problem of Boneh and Lipton the linear 
structure is of certain periodic functions from the additive group 
$\mathbb{Z}^k$ to some arbitrary range.
This does not result in the quantum vs.\ classical query complexity 
separation that we obtain.

It should also be noted that our hidden linear structure problem 
is a special case of the {\em hidden subgroup\/} problem defined 
by Brassard and H{\o}yer \cite{BrassardH97} and Mosca and
Ekert \cite{MoscaE99}.
(This relationship was pointed out to us by Hallgren \cite{Hallgren01}.)
However, using standard techniques for the hidden subgroup problem results 
in a quantum algorithm solving the hidden linear structure problem 
with $\Theta(n)$ queries, as opposed to a single query as required by 
our algorithm.

Finally, one may also consider a variant of our hidden linear structure
problem defined over a finite ring (such as $\mathbb{Z}_{2^n}$) rather
than a field.
However, the exponential classical query complexity lower bound depends
on the field structure and does not always hold for finite rings.
For example, in the case of $\mathbb{Z}_{2^n}$, the classical query 
complexity is $n+1$ rather than exponential (this is explained in 
section~\ref{sec:hidden}).

Our single-query quantum algorithm for the hidden linear structure 
problem is based on an extension of the quantum Fourier transform (QFT) 
to finite fields whose behavior has natural properties with respect to 
the field structure.
This QFT is motivated and defined in section~\ref{sec:qft}, where an 
efficient quantum algorithm for it is also given.
The quantum algorithm and classical lower bound for the hidden linear 
structure problem are given in section~\ref{sec:hidden}.
In section~\ref{sec:matrices}, the QFT is generalized to rings of 
matrices over finite fields.

{\bf Related work.}
Van~Dam and Hallgren have independently proposed a definition for 
QFTs over finite fields that is similar to ours,
and have applied these transforms in the context of black-box problems 
called the ``shifted quadratic character problems''.
Their work first appeared as \cite{vanDamH00} and the preliminary version
of 
this paper appeared as \cite{BCW00}.
 
%============================================================================%

\section{Quantum Fourier transforms for finite fields}
\label{sec:qft}

In this section we propose a definition for quantum Fourier transforms 
over finite fields, whose behavior has natural properties with respect 
to a given field's structure.
We also show how to compute these transformations efficiently.

We assume the reader is familiar with basic concepts regarding finite
fields and computations over finite fields (see, for instance,
\cite{Cohen93, GathenG99, LidlN94}).
As usual, we let $\GF(q)$ denote the finite field having $q = p^n$
elements for some prime $p$.
We assume that an irreducible polynomial
$f(Z) = Z^n - \sum_{j=0}^{n-1}a_j Z^j$ over $\GF(p)$ is fixed, 
and that elements of $\GF(q)$ are represented as polynomials over
$\GF(p)$ modulo $f$ in the usual way.
We will write $x = (x_0,\ldots,x_{n-1})$ to denote the field 
element corresponding to $x_0+x_1 Z+\cdots+x_{n-1}Z^{n-1}$, 
and we identify $x$ with the column vector
$\vec{x} = [x_0,\ldots,x_{n-1}]^{\sf T}$.

\begin{definition}\rm
Let $\phi:\GF(q)\rightarrow \GF(p)$ be any nonzero linear mapping
(viewing elements of $\GF(q)$ as $n$ dimensional vectors over $\GF(p)$ as
above).
Then we define the {\em quantum Fourier transform (QFT) over $\GF(q)$ 
relative to $\phi$} (denoted $F_{q,\phi}$) as follows.
For each $x\in \GF(q)$,
\[
F_{q,\phi}:\ket{x}\mapsto\frac{1}{\sqrt{q}}\sum_{y\in GF(q)}
\omega^{\phi(x y)}\ket{y}
\]
for $\omega = e^{2\pi i/p}$, and let $F_{q,\phi}$ be extended to arbitrary
quantum states by linearity.
\end{definition}

\noindent
A natural choice for $\phi$ is the trace, since this gives a transform
independent of the choice of $f$.
However, we will not require this property, and so we allow $\phi$ to be
arbitrary.
It should be noted that, for any prime $q$, the above Fourier transform 
is essentially identical in form to the conventional cyclic Fourier
transform 
modulo $q$.

\begin{figure}[!ht]
\centering

\setlength{\unitlength}{0.032in}

\begin{picture}(60,30)(0,10)

\put(0,15){\line(1,0){10}}
\put(20,15){\line(1,0){20}}
\put(50,15){\line(1,0){10}}

\put(0,30){\line(1,0){10}}
\put(20,30){\line(1,0){20}}
\put(50,30){\line(1,0){10}}

\put(30,15){\circle{5}}

\put(30,30){\circle*{3}}

\put(30,12.5){\line(0,1){17.5}}
\put(26,21){\makebox(3,3){$s$}}

\put(10,10){\framebox(10,10){$F^{\mbox{\scriptsize \dag}}$}}
\put(10,25){\framebox(10,10){$F$}}
\put(40,10){\framebox(10,10){$F$}}
\put(40,25){\framebox(10,10){$F^{\mbox{\scriptsize \dag}}$}}

\end{picture}
\begin{picture}(14,30)(0,3.0)

\put(0,0){\makebox(14,30){$\equiv$}}

\end{picture}
\begin{picture}(20,30)(0,10)

\put(0,15){\line(1,0){20}}
\put(0,30){\line(1,0){20}}

\put(10,15){\circle*{3}}

\put(10,30){\circle{5}}

\put(10,15){\line(0,1){17.5}}
\put(11.3,21){\makebox(3,3){$s$}}

\end{picture}
\caption{\small The control/target inversion property.}
\label{fig3}
\end{figure}
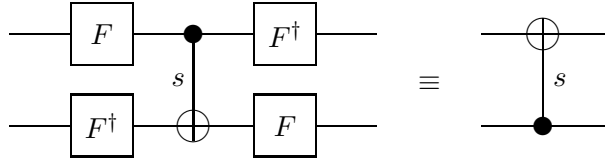

An important property of these transformations is illustrated in
Figure~\ref{fig3}, where $F$ denotes the QFT and the two-register gate
labeled by $s \in \GF(q)$ denotes the mapping 
$\ket{x}\ket{y} \mapsto \ket{x}\ket{y + s x}$.
Let us refer to the latter gate as a {\em controlled-ADD$_s$} gate, 
with its first input called the {\em control\/} register and its 
second input called the {\em target\/} register.
The property illustrated in the figure will be referred to as the
{\em control/target inversion property}.
In words, conjugating a controlled-ADD$_s$ gate by 
$F \otimes F^{\dagger}$ switches its control and target 
registers.
In the special case of $\GF(2)$, $F$ is the Hadamard gate and 
the two-qubit gate is the controlled-NOT gate (when $s=1$).

\begin{theorem}\sl
For $q = p^n$ and any nonzero linear mapping $\phi:\GF(q)\rightarrow
\GF(p)$, 
$F_{q,\phi}$ is unitary and satisfies the control/target inversion
property 
of Figure~\ref{fig3}.
\end{theorem}

\loud{Proof:}
First let us show that $F_{q,\phi}^{\dagger}F_{q,\phi}\ket{x}=\ket{x}$
for every $x\in \GF(q)$.
We have
\begin{multline*}
F_{q,\phi}^{\dagger}F_{q,\phi}\ket{x}
= F_{q,\phi}^{\dagger}\frac{1}{\sqrt{q}}\sum_{y\in GF(q)}
\omega^{\phi(x y)}\ket{y}
= \frac{1}{q}\sum_{y\in GF(q)}\sum_{z \in GF(q)}
\omega^{\phi(x y)-\phi(y z)}\ket{z}\\
= \sum_{z \in GF(q)}\left(\frac{1}{q}\sum_{y\in GF(q)}
\omega^{\phi(y (x-z))}\right)\ket{z}
= \ket{x},
\end{multline*}
following from the fact that $\phi(w)$ must be uniformly
distributed over $\GF(p)$ as $w$ ranges over $\GF(q)$ (since $\phi$ 
is linear and not identically zero).

Next let us verify that the control/target inversion property holds, namely
that for $A_s$ and $B_s$ defined by $A_s\ket{x}\ket{y}=\ket{x}\ket{y + s
x}$
and $B_s\ket{x}\ket{y}=\ket{x + s y}\ket{y}$ we have
\[
(F_{q,\phi}^{\dagger}\otimes F_{q,\phi})A_s
(F_{q,\phi}\otimes F_{q,\phi}^{\dagger})
= B_s.
\]
To prove this relation holds, let us define
\[
\ket{\psi_x} = F_{q,\phi}\ket{x} =
\frac{1}{\sqrt{q}}\sum_{y\in GF(q)}\omega^{\phi(x y)}\ket{y}
\]
for each $x\in \GF(q)$, and note that for $P_w$ defined by
$P_w\ket{x} = \ket{x + w}$ we have
\[
P_w\ket{\psi_{-x}} = 
\frac{1}{\sqrt{q}}\sum_{y\in GF(q)}
\omega^{-\phi(x y)}\ket{y+w}
= 
\frac{1}{\sqrt{q}}\sum_{y\in GF(q)}
\omega^{-\phi(x y - x w)}\ket{y}
=
\omega^{\phi(x w)}\ket{\psi_{-x}}.
\]
Now, for each $x,y\in \GF(q)$ we have
\begin{eqnarray*}
(F_{q,\phi}^{\dagger}\otimes F_{q,\phi})A_s
(F_{q,\phi}\otimes F_{q,\phi}^{\dagger})
\ket{x}\ket{y}
& = &
(F_{q,\phi}^{\dagger}\otimes F_{q,\phi})A_s
\left(\frac{1}{\sqrt{q}}
\sum_{z\in GF(q)}
\omega^{\phi(x z)}\ket{z}\ket{\psi_{-y}}\right)\\
& = &
(F_{q,\phi}^{\dagger}\otimes F_{q,\phi})
\left(\frac{1}{\sqrt{q}}
\sum_{z\in GF(q)}
\omega^{\phi(x z)}\omega^{\phi(y s z)}\ket{z}
\ket{\psi_{-y}}\right)\\
& = &
(F_{q,\phi}^{\dagger}\otimes F_{q,\phi})
\ket{\psi_{x + s y}}\ket{\psi_{-y}}\\[2mm]
& = &
\ket{x + s y}\ket{y}\\[2mm]
& = &
B_{s}\ket{x}\ket{y}
\end{eqnarray*}
as required.
\qed\ee

Next we describe quantum circuits for performing $F_{q,\phi}$ and
analyze their complexity.
Let $C(p,\e)$ denote the minimum size of a quantum circuit
approximating the quantum Fourier transform modulo $p$ to within 
accuracy $\e$.
Note that $C(p,0) \in O(p^2 \log p)$ \cite{BarencoB+95} and, for $\e >
0$,
$C(p,\e) \in O(\log p \log\log p + \log p \log 1/\e)$ when 
$\e \in \Omega(1/p)$ \cite{HalesH00}.

\begin{theorem}\label{circuit_size}\sl
For $q = p^n$ and any nonzero linear mapping $\phi:\GF(q)\rightarrow
\GF(p)$, 
$F_{q,\phi}$ can be performed with accuracy $\varepsilon$ by a quantum
circuit
of size $O(n^2(\log p)^2) + nC(p,\e/n)$.
\end{theorem}

\ni Thus, when $p = 2$ (or any constant), the QFT circuit size is $O(n^2)$
in
the exact case.\ee

\loud{Proof of Theorem~\ref{circuit_size}:}
For any choice of $\phi$ (linear and nonzero), there exists a uniquely
determined $n\times n$ matrix $M_\phi$ over $\GF(p)$ such that
$\phi(xy) = \vec{x}^{\trans}M_\phi\vec{y}$.
We show how to efficiently obtain such a matrix $M_\phi$ explicitly for any
given $\phi$ below.
The quantum circuit performing $F_{q,\phi}$ will depend on $M_\phi$, and
we note that $M_\phi$ must be invertible.

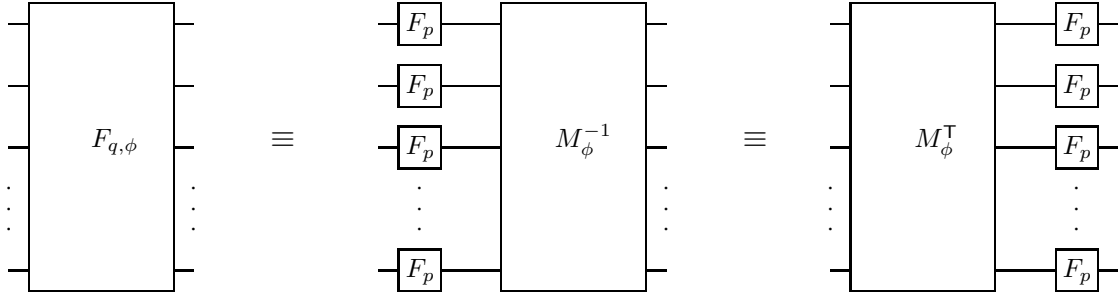
\begin{figure}
\begin{center}
\setlength{\unitlength}{1697sp}%
\begingroup\makeatletter\ifx\SetFigFont\undefined%
\gdef\SetFigFont#1#2#3#4#5{%
  \reset@font\fontsize{#1}{#2pt}%
  \fontfamily{#3}\fontseries{#4}\fontshape{#5}%
  \selectfont}%
\fi\endgroup%
\begin{picture}(16262,4224)(851,-5473)
\thinlines
\put(901,-3961){\circle*{45}}
\put(901,-4261){\circle*{45}}
\put(901,-4561){\circle*{45}}
\put(3601,-3961){\circle*{45}}
\put(3601,-4261){\circle*{45}}
\put(3601,-4561){\circle*{45}}
\put(6901,-3961){\circle*{45}}
\put(6901,-4261){\circle*{45}}
\put(6901,-4561){\circle*{45}}
\put(10501,-3961){\circle*{45}}
\put(10501,-4261){\circle*{45}}
\put(10501,-4561){\circle*{45}}
\put(12901,-3961){\circle*{45}}
\put(12901,-4261){\circle*{45}}
\put(12901,-4561){\circle*{45}}
\put(16501,-3961){\circle*{45}}
\put(16501,-4261){\circle*{45}}
\put(16501,-4561){\circle*{45}}
\put(16201,-1861){\framebox(600,600){}}
\put(901,-1561){\line( 1, 0){300}}
\put(3301,-1561){\line( 1, 0){300}}
\put(12901,-1561){\line( 1, 0){300}}
\put(15301,-1561){\line( 1, 0){900}}
\put(6601,-1861){\framebox(600,600){}}
\put(6301,-1561){\line( 1, 0){300}}
\put(10201,-1561){\line( 1, 0){300}}
\put(7201,-1561){\line( 1, 0){900}}
\put(6601,-2761){\framebox(600,600){}}
\put(6601,-3661){\framebox(600,600){}}
\put(6601,-5461){\framebox(600,600){}}
\put(8101,-5461){\framebox(2100,4200){}}
\put(6301,-2461){\line( 1, 0){300}}
\put(6301,-3361){\line( 1, 0){300}}
\put(6301,-5161){\line( 1, 0){300}}
\put(7201,-2461){\line( 1, 0){900}}
\put(7201,-3361){\line( 1, 0){900}}
\put(7201,-5161){\line( 1, 0){900}}
\put(10201,-2461){\line( 1, 0){300}}
\put(10201,-3361){\line( 1, 0){300}}
\put(10201,-5161){\line( 1, 0){300}}
\put(1201,-5461){\framebox(2100,4200){}}
\put(901,-2461){\line( 1, 0){300}}
\put(3301,-2461){\line( 1, 0){300}}
\put(3301,-5161){\line( 1, 0){300}}
\put(901,-5161){\line( 1, 0){300}}
\put(901,-3361){\line( 1, 0){300}}
\put(3301,-3361){\line( 1, 0){300}}
\put(16801,-1561){\line( 1, 0){300}}
\put(13201,-5461){\framebox(2100,4200){}}
\put(16201,-2761){\framebox(600,600){}}
\put(15301,-2461){\line( 1, 0){900}}
\put(16801,-2461){\line( 1, 0){300}}
\put(12901,-2461){\line( 1, 0){300}}
\put(16201,-3661){\framebox(600,600){}}
\put(15301,-3361){\line( 1, 0){900}}
\put(16801,-3361){\line( 1, 0){300}}
\put(12901,-3361){\line( 1, 0){300}}
\put(12901,-5161){\line( 1, 0){300}}
\put(15301,-5161){\line( 1, 0){900}}
\put(16201,-5461){\framebox(600,600){}}
\put(16801,-5161){\line( 1, 0){300}}
\put(2101,-3361){\makebox(0,0)[lb]
{\smash{\SetFigFont{10}{13.2}{\rmdefault}{\mddefault}{\updefault}$F_{q,\phi}$}}}
\put(8751,-3361){\makebox(0,0)[lb]
{\smash{\SetFigFont{10}{13.2}{\rmdefault}{\mddefault}{\updefault}
$M_\phi^{-1}$}}}
\put(14001,-3361){\makebox(0,0)[lb]
{\smash{\SetFigFont{10}{13.2}{\rmdefault}{\mddefault}{\updefault}
$M^{\trans}_\phi$}}}
\put(6701,-1661){\makebox(0,0)[lb]
{\smash{\SetFigFont{10}{13.2}{\rmdefault}{\mddefault}{\updefault}$F_p$}}}
\put(6701,-2561){\makebox(0,0)[lb]
{\smash{\SetFigFont{10}{13.2}{\rmdefault}{\mddefault}{\updefault}$F_p$}}}
\put(6701,-3461){\makebox(0,0)[lb]
{\smash{\SetFigFont{10}{13.2}{\rmdefault}{\mddefault}{\updefault}$F_p$}}}
\put(6701,-5261){\makebox(0,0)[lb]
{\smash{\SetFigFont{10}{13.2}{\rmdefault}{\mddefault}{\updefault}$F_p$}}}
\put(16301,-1661){\makebox(0,0)[lb]
{\smash{\SetFigFont{10}{13.2}{\rmdefault}{\mddefault}{\updefault}$F_p$}}}
\put(16301,-2561){\makebox(0,0)[lb]
{\smash{\SetFigFont{10}{13.2}{\rmdefault}{\mddefault}{\updefault}$F_p$}}}
\put(16301,-3461){\makebox(0,0)[lb]
{\smash{\SetFigFont{10}{13.2}{\rmdefault}{\mddefault}{\updefault}$F_p$}}}
\put(16301,-5261){\makebox(0,0)[lb]
{\smash{\SetFigFont{10}{13.2}{\rmdefault}{\mddefault}{\updefault}$F_p$}}}
\put(11626,-3361){\makebox(0,0)[lb]
{\smash{\SetFigFont{10}{13.2}{\rmdefault}{\mddefault}{\updefault}{\large
$\equiv$}}}}
\put(4726,-3361){\makebox(0,0)[lb]{\smash{\SetFigFont{10}{13.2}{\rmdefault}{\mddefault}{\updefault}{\large
$\equiv$}}}}
\end{picture}
\end{center}
\caption{Equivalent circuits for $F_{q,\phi}$}
\label{fig:circuits1}
\end{figure}

We have
\[
F_{q,\phi}\ket{x} = \frac{1}{\sqrt{q}}\sum_{y\in GF(q)}
\omega^{\vec{x}^{\trans}M_\phi\vec{y}}\ket{y}
= \frac{1}{\sqrt{q}}\sum_{y\in GF(q)}
\omega^{\vec{x}^{\trans}\vec{y}}\ket{M_\phi^{-1}\vec{y}}
= \frac{1}{\sqrt{q}}\sum_{y\in GF(q)}
\omega^{(M_\phi^{\trans}\vec{x})^{\trans}\vec{y}}
\ket{y}.
\]
From this we conclude that
\[
F_{q,\phi} = M_\phi^{-1}(F_p\otimes\cdots\otimes F_p)
= (F_p\otimes\cdots\otimes F_p)M_\phi^{\trans},
\]
where $F_p$ denotes the usual quantum Fourier transform modulo $p$
and, for $A\in\{M_\phi^{-1}, M_\phi^{\trans}\}$, we identify $A$
with the reversible operation that maps each $\ket{\vec{x}}$ to
$\ket{A\vec{x}}$.
This relation is illustrated in Figure~\ref{fig:circuits1}.

The upper bound of $O(n^2(\log p)^2) + nC(p,\varepsilon/n)$ now follows
from
the observation that in order to implement $F_{q,\phi}$ with accuracy
$\varepsilon$ it suffices to implement each circuit for $F_p$ with accuracy
$\varepsilon/n$ (contributing $n C(p,\varepsilon/n)$ gates to the final
circuit) and to implement the circuit for multiplication by either
$M_\phi^{\trans}$ or $M_\phi^{-1}$ exactly.
Let $A\in\{M_\phi^{-1}, M_\phi^{\trans}\}$.
Multiplication of an $n$-dimensional vector $v$ by $A$ can be done
with $O(n^2)$ arithmetic operations in $\GF(p)$, each of which can be
performed by a circuit of size $O((\log p)^2)$, resulting in a circuit
of size $O(n^2(\log p)^2)$.
In order to implement this transformation reversibly within the same size
bound, it suffices to be able to invert the computation in this size bound.
Inverting this computation is simply multiplication by $A^{-1}$,
which can be performed in precisely the same size bound.
(Note that the circuit itself does not need to invert $A$, but
rather information about $A$ and $A^{-1}$ is pre-computed and
``hard-coded'' into the appropriate circuit for $F_{q,\phi}$.)

Now let us return to the question of determining the matrix $M_\phi$
corresponding to a given $\phi$.
First, note that multiplication of field elements satisfies
\[
(z_0,\ldots,z_{n-1}) = (x_0,\ldots,x_{n-1})\cdot (y_0,\ldots,y_{n-1})
\]
where
\begin{equation}
z_i = \vec{x}^{\trans} B_i\vec{y}
\label{eq:B_matrices}
\end{equation}
for a certain sequence of $n\times n$ matrices $B_0,\ldots,B_{n-1}$
over $\GF(p)$.

Let us explicitly construct a sequence $B_0,\ldots,B_{n-1}$ that satisfies
Eq.~\ref{eq:B_matrices}.
To do this, it will be helpful to review the notion of {\em Hankel
matrices}.
An $n\times n$ Hankel matrix $A$ is a matrix of the form
\begin{equation}
A = \left[
\begin{array}{lllcl}
t_0     & t_1     & t_2     & \cdots & t_{n-1} \\[2mm]
t_1     & t_2     & t_3     & \cdots & t_n \\[2mm]
t_2     & t_3     & t_4     & \cdots & t_{n+1} \\
\vdots  & \vdots  & \vdots  & \ddots & \vdots \\
t_{n-1} & t_n\;\; & t_{n+1} & \cdots & t_{2n-2}
\end{array}
\right].
\label{eq:Hankel}
\end{equation}
That is, the ``anti-diagonals'' each contain only one element (or,
equivalently,
$A[i,j]$ depends only on $i+j$).
The Hankel matrix in Eq.~\ref{eq:Hankel} will be denoted
$\op{Hankel}(t_0,t_1,\ldots,t_{2n-2})$.

Recall that we have
\[
Z^n\:\equiv\:\sum_{j=0}^{n-1}a_j Z^j \;\;\;\;(\bmod\,f(Z)),
\]
where $f$ is as described at the beginning of the current section.
Write $a_j^{(0)} = a_j$ for $j = 0,\ldots,n-1$.
We will actually need numbers $a_j^{(k)}$ (for $j = 0,\ldots, n-1$,
$k = 0,\ldots,n-2$) such that
\[
Z^{n+k}\:\equiv\:\sum_{j=0}^{n-1}a_j^{(k)} Z^j \;\;\;\;(\bmod\,f(Z)).
\]
These numbers are easy to obtain.
Define an $n\times n$ matrix $V$ as follows:
\[
V = \left[
\begin{array}{ccccl}
0 & 0 & \cdots & 0 & a_0 \\
1 & 0 & \cdots & 0 & a_1 \\
0 & 1 & \cdots & 0 & a_2 \\
\vdots & \vdots & \ddots & \vdots & \,\vdots \\
0 & 0 & \cdots & 1 & a_{n-1}
\end{array}\!\!
\right]
\]
Then
\[
\left[a_0^{(k)},\ldots,a_{n-1}^{(k)}\right]^{\trans}
\:=\: V^k \left[a_0,\ldots,a_{n-1}\right]^{\trans}
\:=\: V^{k+1} \left[0,\ldots,0,1\right]^{\trans}.
\]
Finally, we can describe the matrices $B_0,\ldots,B_{n-1}$.
For each $i = 0,\ldots,n-1$,
\[
B_i = \op{Hankel}\left(\delta_{0,i},\delta_{1,i},\ldots,\delta_{n-1,i},
a_i^{(0)},a_i^{(1)},\ldots,a_i^{(n-2)}\right).
\]
(Here, $\delta_{i,j}$ is the Kronecker-$\delta$ symbol.)
A straightforward computation reveals that this choice for
$B_0,\ldots,B_{n-1}$ satisfies Eq.~\ref{eq:B_matrices}.
It is also not hard to show that these matrices $B_0,\ldots,B_{n-1}$ are
the only matrices satisfying Eq.~\ref{eq:B_matrices}, and that each
$B_i$ is
necessarily invertible.

Now, since $\phi:\GF(q)\rightarrow \GF(p)$ is linear and not identically
zero,
we must have $\phi(x) = \sum_{i=0}^{n-1}\lambda_i x_i$
for each $x\in\GF(q)$ for some choice of
$\lambda_0,\ldots,\lambda_{n-1}\in\GF(p)$ (not all 0).
At this point we see that
$\phi(x y) = \vec{x}^{\trans}M_\phi\vec{y}$
for $M_\phi = \sum_{i=0}^{n-1} \lambda_i B_i$.
Equivalently, we have
\[
M_\phi = \op{Hankel}\left(\lambda_0,\ldots,\lambda_{n-1},
\sum_{i=0}^{n-1}\lambda_i a_i^{(0)},\ldots,
\sum_{i=0}^{n-1}\lambda_i a_i^{(n-2)}\right).
\]
\qed

In the previous theorem, we have ignored the issue of circuit uniformity.
However, it follows from the proof that each circuit for $F_{q,\phi}$ can
be generated in polynomial time under a similar assumption on the circuits
for performing $F_p$.

%============================================================================%

\section{The hidden linear structure problem}
\label{sec:hidden}

For a prime power $q$, define the {\em hidden linear structure\/} problem 
over $\GF(q)$ as follows.
In the classical version, one is given a 
black-box that maps 
$(x,y) \in \GF(q) \times \GF(q)$ to $(x,\pi(y + s x))$, where $\pi$ is an 
arbitrary permutation on the elements of $\GF(q)$ and $s \in \GF(q)$.
Analogously, in the quantum case, one is given a black-box performing 
the unitary transformation that maps $\ket{x}\ket{y}$ 
($x, y \in \GF(q)$) to $\ket{x}\ket{\pi(y + s x)}$.
The goal is to determine the value of $s$.

In this section, we give a sharp quantum vs.\ classical query complexity 
separation for the hidden linear structure problem.
First, in the classical case, $\Omega(\sqrt{q})$ queries are necessary to 
solve this problem, even with bounded error.
Second, in the quantum case, a single quantum query is sufficient to solve 
the hidden linear structure problem exactly, provided that one can compute 
the QFTs $F_{q,\phi}$ and $F_{q,\phi}^{\dagger}$.
In the case where $q=2^n$, the QFT can be performed exactly with only 
$O(n^2)$ basic operations (Hadamard gates and controlled-NOT gates).
The result is a single-query exact quantum algorithm to extract $s$ 
with $O(n^2)$ auxiliary operations.
Moreover, in this case the algorithm can be streamlined so as to consist 
of $O(n)$ Hadamard gates, the single query, and $O(n^2)$ classical 
post-processing after a measurement is made.
In the case where $q$ is an $n$-bit prime, our results are weaker, since 
the best procedure that we are aware for performing the QFT exactly in that
case is $O(p^2 \log p) = O(n 4^n)$.

It should be noted that if the finite fields are relaxed to 
{\em finite rings\/} then, for the analogous hidden linear structure
problem, 
the quantum vs.\ classical classical query complexity separation may be
much 
weaker.
This is because the classical query complexity of the problem can become 
much smaller.
For example, for the ring ${\mathbb Z}_{2^n}$, there is a simple classical 
procedure solving the hidden linear structure problem with only $n+1$
queries.
It begins by querying $(0,0)$ and $(2^{n-1},0)$, yielding $\pi(0)$ and 
$\pi(s 2^{n-1})$ respectively.
If $\pi(0) = \pi(s 2^{n-1})$ then $s$ is even; otherwise $s$ is odd.
Thus, two queries reduce the number of possibilities for $s$ by a factor 
of 2.
If $s$ is even then the next query is $(2^{n-2},0)$, yielding 
$\pi(s 2^{n-2})$, which determines whether $s \bmod 4$ 
is 0 or 2.
If $s$ is odd then the next query is $(2^{n-2},2^n-2^{n-2})$, 
yielding $\pi(2^n - 2^{n-2} + s 2^{n-2})$, which determines 
whether $s \bmod 4$ is 1 or 3.
This process can be continued so as to deduce $s$ after $n+1$ 
queries.
For this reason, our attention is focused on the hidden linear structure
problem over fields (though we do consider QFTs for some noncommutative
rings
in the next section).

We proceed with the classical lower bound.

\begin{theorem}\label{structure-lb}\sl 
$\Omega(\sqrt{q})$ queries are necessary to solve the hidden linear
structure
problem over $\GF(q)$ within error probability $\frac{1}{2}$.
\end{theorem}

\loud{Proof:}
The lower bound proof is similar to that for Simon's problem
\cite{Simon94}.
First, by a game-theoretic argument \cite{Yao83}, it suffices to consider 
deterministic algorithms where the input data, embodied by the values of 
$s$ and $\pi$, is probabilistic.
Set both $s \in \GF(q)$ and $\pi$ (a permutation on $\GF(q)$) randomly, 
according to the uniform distribution.
Consider the information obtained about $s$ after $k$ queries 
$(x_1,y_1),\ldots,(x_k,y_k)$ (without loss of generality, the queries 
are all distinct).
If, for some $i \neq j$, the outputs of the $i^{\mbox{\scriptsize th}}$ 
and $j^{\mbox{\scriptsize th}}$ queries collide in that
$\pi(y_i + s x_i) = \pi(y_j + s x_j)$, then $y_i + s x_i = y_j + s x_j$,
which implies that the value of $r$ can be determined as
\begin{equation}
s = \frac{y_i - y_j}{x_j - x_i}
\end{equation}
(note that $x_j - x_i \neq 0$, since this would imply that 
$(x_i,y_i) = (x_j,y_j)$).
On the other hand, if there are no collisions among the outputs of all $k$
queries then all that can be deduced about $s$ is that 
\begin{equation}
s \neq \frac{y_i - y_j}{x_j - x_i}
\end{equation}
for all $1 \le i < j \le k$.
This leaves $q - k(k-1)/2$ values for $s$, which are equally likely
by symmetry.

Now, consider the probability of a collision occurring at the 
$k^{\mbox{\rm\scriptsize th}}$ query given that no collisions have 
occurred in the previous $k-1$ queries.
After the first $k-1$ queries, there remain at least 
$q - (k-1)(k-2)/2 > q - k^2/2$ possible values of $s$, equally likely
by symmetry.
Of these values, at most $k-1$ induce a collision between the 
$k^{\mbox{\rm\scriptsize th}}$ query and one of the $k-1$ previous 
queries.
Therefore, the probability of a collision occurring at the 
$k^{\mbox{\rm\scriptsize th}}$ query is at most 
\begin{equation}
\frac{k-1}{q - k^2/2} \le \frac{2k}{2q - k^2}.
\end{equation}
It follows that the probability of a collision occurring at all during 
the first $l$ queries is bounded above by 
\begin{equation}
\sum_{k=1}^{l}\frac{2k}{2q - k^2} \le 
\frac{l^2}{2q - l^2}.
\end{equation}
If this probability is to be greater than or equal to $1/2$ then
$l^2/(2q - l^2) \ge 1/2$, which implies that 
\begin{equation}
l \ge \sqrt{2q/3} \in \Omega(\sqrt{q}).
\end{equation}
\qed\ee

Next, we describe the quantum algorithm.

\begin{theorem}\sl
For a given field $\GF(q)$, if $F_{q,\phi}$ and $F^{\dag}_{q,\phi}$ can
be performed for some nonzero linear mapping $\phi$ then a single query 
is sufficient to solve the hidden linear structure problem exactly.
\end{theorem}

\loud{Proof:}\label{structure-ub}
The quantum procedure is to initialize the state of two $\GF(q)$-valued 
registers to $\ket{0}\ket{1}$ (where 0 and 1 are respectively the 
additive and multiplicative identities of the field) and perform the 
following operations (where $F = F_{q,\phi}$):
\begin{enumerate}
\item Apply $F \otimes F^{\dag}$.
\item Query the black box.
\item Apply $F^{\dag} \otimes F$.
\end{enumerate}
Then the state of the first register is measured.

Tracing through the evolution of the state of the registers 
during the execution of the above algorithm, the state after 
each step is:
\begin{enumerate}
\item $(F\ket{0})(F^{\dag}\ket{1})$
\item $(F\ket{s})(U_{\pi} F^{\dag}\ket{1})$
\item $\ket{s}(F U_{\pi} F^{\dag}\ket{1})$
\end{enumerate}
The transformation from step 1 to step 2 follows from the control/target 
inversion property, as shown in figure~\ref{fig3}.
It is clear that the output of the algorithm is $s$.
\qed\ee

As mentioned previously, the transformation $F_{2^n,\phi}$ for any
$\phi$ is particularly simple, and yields the following algorithm.

\begin{enumerate}
\item Initialize the state of two $\GF(2^n)$-valued registers to
the (classical) state $\ket{0}\ket{M_{\phi}\vec{1}}$.
\item
Apply a Hadamard transform to each qubit of each register.
\item
Query the black-box.
\item
Apply a Hadamard transform to each qubit of each register.
\item
Measure the first register, yielding an $n$-bit string $z$.
\item
Classically, compute $(M_{\phi}^{\trans})^{-1}\vec{z}$.
\end{enumerate}
The result will be $s$.

%=============================================================================%

\section{Extension to Rings}
\label{sec:matrices}

It is natural to generalize the concept of controlled addition as we have
seen
it to rings in general.
So, one might ask whether, for all rings, there exist operations
corresponding
to ``quantum Fourier transforms'' in the sense that they perform
control/target inversion on controlled-addition gates over that ring.
While we do not know the answer to this question, we will show that for any
commutative ring $R$ where such a Fourier transform exists, it is possible
to
define quantum Fourier transforms for the noncommutative ring of $\sqx{m}$
matrices over $R$.

Let us introduce some notation. In this section, all matrices are
understood
to be square matrices.
Given an $m^2$ array of quantum registers $\{E_{ij}\}$ over a commutative
ring
$R$, we associate the state $\ket{x_{ij}}$ with the register $E_{ij}$.
We also identify the $\sqx{m}$ matrix $X$ given by
\[
        X =\left[\begin{array}{cccc}
        x_{11} & x_{12} & \ldots & x_{1m} \\
        x_{21} & x_{22} & \ldots & x_{2m} \\
        \vdots & \vdots & \ddots & \vdots \\
        x_{m1} & x_{m2} & \ldots & x_{mm} \end{array}\right]
\]
with the product state
\[
        \ket{X} = \bigotimes_{i=1}^{m} \bigotimes_{j=1}^{m} \ket{x_{ij}}
        = \ket{x_{11}} \ket{x_{12}} \ldots \ket{x_{1m}}
          \ket{x_{21}} \ldots \ket{x_{mm}}
\]
of the states of the registers. We then make the following definition.
\label{QFTmat}\begin{definition}\rm
Let $F_R$ be a quantum Fourier transform over a commutative ring $R$.
Then we define the quantum Fourier transform over $\sqmtx{m}{R}$ by the
following mapping for each matrix $X = (x_{ij}) \in \sqmtx{m}{R}$:
\[
        F_{R,m}: \ket{X} \mapsto
        \bigotimes_{i=1}^{m} \bigotimes_{j=1}^{m} F_R \ket{x_{ji}}.
\]
That is, the quantum Fourier transform of $\ket{X}$ is performed by
applying
the Fourier transform $F_R$ independently to all the quantum registers used
to
represent $X$, and transposing those registers (or their states) within the
register array.
\end{definition}

Multiplication in matrix rings over $R$ will, in general, be
non-commutative.
Therefore, in working with matrices, we must distinguish between left and
right multiplication when defining the controlled addition operators.
We define left-controlled addition with parameter $S$ (denoted by
\lcadd{S})
and right-controlled addition with parameter $S$ (denoted by \rcadd{S}) by
the
following action on basis states:
\[
        \begin{array}{lcl}
        \lcadd{S}: \ket{X}\ket{Y} \mapsto \ket{X}\ket{Y + S X}
        & &
        \rcadd{S}: \ket{X}\ket{Y} \mapsto \ket{X}\ket{Y + X S}
        \end{array}
\]
As well, we introduce left and right controlled addition operators with the
roles of the target and control registers reversed:
\[
        \begin{array}{lcl}
        \rlcadd{S}: \ket{X}\ket{Y} \mapsto \ket{X + S Y}\ket{Y}
        & &
        \rrcadd{S}: \ket{X}\ket{Y} \mapsto \ket{X + Y S}\ket{Y}
        \end{array}
\]
As the order of multiplication becomes important for rings in general, we
find
it reasonable to make the following expansion of the definition of
control/target inversion: a gate $G$ performs control/target inversion on
controlled addition gates over a given ring if the following equality
holds:
\[
        (G^\dagger \ox G) \lcadd{S} (G \ox G^\dagger) = \rrcadd{S}.
\]
That is, in addition to the roles of target and control being interchanged,
the
manner of multiplication (left or right) is switched. In the case where the
ring
is commutative, this reduces to the definition given previously (see
Figure~\ref{fig3}). We will now show that the quantum Fourier
transform $F_{R,m}$ defined above has this property for $\sqx{m}$ matrices
over $R$, when $F_R$ is defined and has the control/target inversion
property
on $R$.

For input matrices $X$ and $Y$ over $R$, we denote 
\[ 
        \ket{X} = \bigotimes_{i=1}^{m}\bigotimes_{j=1}^{m}\ket{x_{ij}}
        \mathtab \mathtab \mathtab
        \ket{Y} = \bigotimes_{i=1}^{m}\bigotimes_{j=1}^{m}\ket{y_{ij}}
\]

Let $E_{ij}$ represent the register which stores the state $\ket{x_{ij}}$,
and
$F_{ij}$ represent the register which stores the state $\ket{y_{ij}}$.
Define the operator $A_{ik}^{ij}(s)$ as a controlled-ADD$_s$ gate which
operates on a control register $E_{ik}$ and a target register $F_{ij}$, and
$B_{ik}^{ij}(s)$ as a controlled-ADD$_s$ gate which operates on a control
register $F_{ik}$ and a target register $E_{ij}$.
Then we can decompose $\lcadd{S}$ as the following product of operators:
\[
        \lcadd{S} =
        \prod_{i=1}^m\prod_{j=1}^m\prod_{k=1}^m A_{ij}^{ik} (s_{kj})
\]
This can be easily verified by testing the effect of this product on the
$ij$-th
target register, where we see that the effect (for basis states) is to add 
the term $x_{ik}s_{kj}$ for each $1 \leq k \leq m$. Control/target
inversion is
expressed for these gates in the following manner:
\[
        ({{F_R}^\dagger}^{\ox m^2} \ox {F_R}^{\ox m^2})
        A_{ij}^{ik}(s_{kj})
        ({F_R}^{\ox m^2} \ox {{F_R}^\dagger}^{\ox m^2})
        = B_{ij}^{ik}(s_{kj}).
\]
Here, the quantum Fourier transforms cancel one another out on all
registers
except the $ij$-th target register and the $ik$-th control register, where
control/target inversion occurs.

Using this decomposition, and applying quantum Fourier transforms to the
individual registers before and after this product of gates in the same
manner
as above, we obtain:
\begin{multline*}
({{F_R}^\dagger}^{\ox m^2} \ox {F_R}^{\ox m^2})\lcadd{S}
({F_R}^{\ox m^2} \ox {{F_R}^\dagger}^{\ox m^2})\\
= \prod_{i=1}^m\prod_{j=1}^m\prod_{k=1}^m B_{ij}^{ik}(s_{kj})
= \prod_{i=1}^m\prod_{k=1}^m\prod_{j=1}^m B_{ij}^{ik}(s_{kj})
= \prod_{i=1}^m\prod_{j=1}^m\prod_{k=1}^m B_{ik}^{ij}(s_{jk})
= \rlcadd{S^{\trans}}
\end{multline*}
That is, the roles of the control and target registers are reversed, and
although the manner of multiplication is unchanged, the parameter matrix
$S$
is transposed.

Note that the quantum Fourier transform $F_{R,m}$ on $\sqx{m}$ matrices
over
$R$ can be decomposed into an application of $F_R$ on each element of the
matrix, and transposing the matrix (denoted by the operator $T_m$), in any
order:
\[
        F_{R,m} = ({F_R}^{\ox m^2}) \trop{m} = \trop{m} ({F_R}^{\ox m^2}).
\]
Clearly, $\trop{m}\trop{m} = I_{m}$ (the identity $\sqx{m}$ matrix).
Then, we can verify that $F_{R,m}$ performs control/target inversion on
controlled addition gates over $\sqmtx{m}{R}$:

\[
        \begin{array}{l}
        (F_{R,m}^\dagger \ox F_{R,m})\lcadd{S}
        (F_{R,m} \ox F_{R,m}^\dagger)\ket{X}\ket{Y}
        \\ \\
        \begin{array}{lll}
        & & = (\trop{m} \ox \trop{m})
        ({F_R^\dagger}^{\ox m^2}\ox {F_R}^{\ox m^2})
        \lcadd{S}
        ({F_R}^{\ox m^2}\ox {F_R^\dagger}^{\ox m^2})
        (\trop{m} \ox \trop{m}) \ket{X}\ket{Y} \\ \\
        & & =
        (\trop{m} \ox \trop{m})
        \rlcadd{S^{\trans}}
        \ket{X^{\trans}}\ket{Y^{\trans}} \\ \\
        & & =
        (\trop{m} \ox \trop{m})
        \ket{X^{\trans} + S^{\trans} Y^{\trans}}\ket{Y^{\trans}} \\ \\ 
        & & = \ket{X + YS}\ket{Y} \\ \\
        & & = \rrcadd{S}\ket{X}\ket{Y},
        \end{array}
        \end{array}
\]
which is what we wished to show.

As for extending the hidden linear structure problem to arbitrary rings, it
is
not clear for which rings $R$ an exponential separation can be achieved.
The ability to perform control/target inversion for this problem when
$R = \sqmtx{m}{\GF(p^n)}$ (for example) indicates that the problem can be
solved in one query in the quantum case, but we do not have strong
classical lower bounds for this case.
However, there do exist rings, such as $\sqx{\GF(p^n)}$, where exponential
separation can be shown, building on the proof for $\GF(p^n)$; thus, the
strong separation in the case of finite fields is not an isolated case.
Considering the proofs of the classical upper bound for $\Z_{p^n}$ and
lower
bound for $\GF(p^n)$, it seems plausible that rings exhibiting a strong
separation will have very few zero divisors, or little additive
structure among the zero divisors.
Both of these statements hold for $\sqx{\GF(p^n)}$, which has a ratio of
$O(1/p^n)$ zero divisors among its elements, and which only has
two ideals which have only a trivial intersection.

%============================================================================%

\section*{Acknowledgments}

R.C. gratefully acknowledges the University of California at Berkeley 
and the California Institute of Technology where some of the writing 
and revisions to this paper occurred.

\bibliographystyle{plain}

\begin{thebibliography}{10}

\bibitem{BarencoB+95}
A.~Barenco, C.~H. Bennett, R.~Cleve, D.~DiVincenzo, N.~Margolus, P.~Shor,
  T.~Sleator, J.~Smolin, and H.~Weinfurter.
\newblock Elementary gates for quantum computation.
\newblock {\em Physical Review A}, 52:3457--3467, 1995.

\bibitem{BCW00}
J.~N~de~Beaudrap, R.~Cleve, J.~Watrous, 
\newblock Quantum Fourier transforms for extracting hidden linear 
structures in finite fields.
\newblock Los Alamos Preprint Archive quant-ph/0011065, 2000.

\bibitem{BernsteinV97}
E.~Bernstein and U.~Vazirani.
\newblock Quantum complexity theory.
\newblock {\em SIAM Journal on Computing}, 26(5):1411--1473, 1997.

\bibitem{BonehL95}
D.~Boneh and R.~Lipton.
\newblock Quantum cryptanalysis of hidden linear functions.
\newblock In {\em Advances in Cryptology -- Crypto'95}, volume 963 of {\em
  Lecture Notes in Computer Science}, pages 242--437. Springer-Verlag,
1995.

\bibitem{BrassardH97}
G.~Brassard and P.~H{\o}yer.
\newblock An exact quantum polynomial-time algorithm for {S}imon's problem.
\newblock In {\em Fifth Israeli Symposium on Theory of Computing and
Systems},
  pages 12--23, 1997.

\bibitem{Cleve00}
R.~Cleve.
\newblock The query complexity of order-finding.
\newblock In {\em Proceedings of the 15th Annual IEEE Conference on
  Computational Complexity}, pages 54--59, 2000.

\bibitem{CleveE+98}
R.~Cleve, A.~Ekert, C.~Macchiavello, and M.~Mosca.
\newblock Quantum algorithms revisited.
\newblock {\em Proceedings of the Royal Society, London}, A454:339--354,
1998.

\bibitem{Cohen93}
H.~Cohen.
\newblock {\em A Course in Computational Algebraic Number Theory}.
\newblock Springer-Verlag, 1993.

\bibitem{vanDamH00}
W.~van Dam and S.~Hallgren.
\newblock Efficient quantum algorithms for shifted quadratic character 
  problems.
\newblock Los Alamos Preprint Archive quant-ph/0011067, 2000.

\bibitem{Deutsch85}
D.~Deutsch.
\newblock Quantum theory, the {Church}--{Turing} principle and the
universal
  quantum computer.
\newblock {\em Proceedings of the Royal Society of London}, A400:97--117,
1985.

\bibitem{DeutschJ92}
D.~Deutsch and R.~Jozsa.
\newblock Rapid solutions of problems by quantum computation.
\newblock {\em Proceedings of the Royal Society of London}, A439:553--558,
  1992.

\bibitem{GathenG99}
J.~von~zur Gathen and J.~Gerhard.
\newblock {\em Modern Computer Algebra}.
\newblock Cambridge University Press, 1999.

\bibitem{HalesH00}
L.~Hales and S.~Hallgren.
\newblock An improved quantum {F}ourier transform algorithm and
applications.
\newblock In {\em Proceedings of the 41st Annual Symposium on Foundations
of 
  Computer Science}, pages 515-525, 2000.

\bibitem{Hallgren01}
S.~Hallgren.
\newblock Personal communication, 2001.

\bibitem{Lam91}
T.Y.~Lam.
\newblock {\em A First Course in Noncommutative Rings}.
\newblock Springer-Verlag, 1991.

\bibitem{LidlN94}
R.~Lidl and H.~Niederreiter.
\newblock {\em Introduction to Finite Fields and Their Applications}.
\newblock Cambridge University Press, revised edition, 1994.

\bibitem{MoscaE99}
M.~Mosca and A.~Ekert.
\newblock The hidden subgroup problem and eigenvalue estimation on a
quantum
  computer.
\newblock In {\em Proceedings of the 1st NASA International Conference on
  Quantum Computing and Quantum Communication}, volume 1509 of {\em Lecture
  Notes in Computer Science}, 1999.
\newblock Also available from the Los Alamos Preprint Archive,
  quant-ph/9903071.

\bibitem{Shor97}
P.~Shor.
\newblock Polynomial-time algorithms for prime factorization and discrete
  logarithms on a quantum computer.
\newblock {\em SIAM Journal on Computing}, 26(5):1484--1509, 1997.

\bibitem{Simon94}
D.~Simon.
\newblock On the power of quantum computation.
\newblock {\em SIAM Journal on Computing}, 26(5):1474--1483, 1997.

\bibitem{Yao83}
A.~C.-C.~Yao.
\newblock Lower bounds by probabilistic arguments.
\newblock In {\em Proceedings of the 24th Annual Symposium on 
  Foundations of Computer Science}, pages 420--428, 1983.

\end{thebibliography}

%============================================================================%

\end{document}